\documentclass[runningheads]{llncs}

\usepackage{graphicx}
\usepackage{tabularx}
\usepackage{float}
\usepackage{booktabs}
\usepackage{adjustbox}
\usepackage{subcaption}
\usepackage{caption}
\usepackage{mwe}

\begin{document}
\title{Exploring the Emotional and Mental Well-Being of Individuals with Long COVID Through Twitter Analysis}
\titlerunning{Long COVID Twitter Analysis}

\author{Guocheng Feng\and
Huaiyu Cai\and
Wei Quan
}
\authorrunning{G. Feng et al.}
%
\institute{Department of Statistics and Data Science, BNU-HKBU United International College, Zhuhai 519087, China \\
\email{\{q030026233,q030026003\}@mail.uic.edu.cn, weiquan@uic.edu.cn}
}

\maketitle              

\begin{abstract}
The COVID-19 pandemic has led to the emergence of Long COVID, a cluster of symptoms that persist after infection. Long COVID patients may also experience mental health challenges, making it essential to understand individuals' emotional and mental well-being. This study aims to gain a deeper understanding of Long COVID individuals' emotional and mental well-being, identify the topics that most concern them, and explore potential correlations between their emotions and social media activity. Specifically, we classify tweets into four categories based on the content, detect the presence of six basic emotions, and extract prevalent topics. Our analyses reveal that negative emotions dominated throughout the study period, with two peaks during critical periods, such as the outbreak of new COVID variants. The findings of this study have implications for policy and measures for addressing the mental health challenges of individuals with Long COVID and provide a foundation for future work.

\end{abstract}

\section{Introduction}
The COVID-19 pandemic has had a profound and diverse impact on individuals and communities across the globe. While numerous individuals have recovered from COVID-19 infection, many patients have reported experiencing chronic symptoms following their acute illness. Since May 2020, individuals experiencing persistent symptoms after contracting COVID-19 have been using Long Hauler to refer to themselves and Long COVID to describe their condition on social media \cite{Callard2020HowAW}. Long COVID refers to a cluster of symptoms that emerge during or after a confirmed or suspected case of COVID-19 \cite{Davis2020CharacterizingLC}. Current research suggests that over 200 symptoms associated with Long COVID have been identified. A conservative estimate indicates that these symptoms afflict 10\% of the infected population and that the number of cases is rising steadily \cite{Ballering2022PersistenceOS}.

Additionally, a study has revealed that a considerable percentage of individuals diagnosed with COVID-19 have reported experiencing mental health disorders such as anxiety, depression, and post-traumatic stress disorder within three to six months of the onset of COVID-19 \cite{HoubenWilke2022TheIO}. Individuals experiencing symptoms associated with Long COVID may also encounter a range of factors that can adversely affect their mental well-being, including symptoms that hinder their daily activities \cite{Burton2022FactorsST}. Hence, it is essential to establish a comprehensive understanding of the mental health challenges that are linked to Long COVID.

The use of social media during the pandemic has provided people with a platform to share opinions and personal experiences, which could be a valuable source of information for researchers seeking early insights into health-related topics. There has been a significant 14\% rise in the manifestation of mental health symptoms on social media platforms during the COVID-19 crisis \cite{Saha2020PsychosocialEO}. 

In this study, we investigate the mental health status of Long COVID Twitter users by analyzing publicly available Twitter data. Specifically, we classify tweets into four categories (Non-Long COVID Tweets, Long COVID Tweets without Mental Health Content, Long COVID Tweets with Implicit Mental Health Statements, and Long COVID Tweets with Explicit Mental Health Statements), detect the presence of Ekman's six basic emotions (Anger, Disgust, Fear, Joy, Sadness, and Surprise) \cite{Ekman1992AnAF} in tweets, and extract topics from each category. We aim to answer the following research questions:
\begin{itemize}
    \item RQ1. What are the distributions of emotions of different categories, and how do these emotions change over time?
    \item RQ2. What are the correlations between the emotions of different categories?
    \item RQ3. What are the prevalent topics in different categories?
\end{itemize}

Our contributions include gaining a better understanding of the emotional and mental well-being of Long COVID Twitter users, identifying the topics that concern them the most, and establishing a potential correlation between their emotions and social media activities.

\section{Background}
Twitter has become a valuable source for the medical community to identify and analyze medical and clinical indicators of Long COVID, enabling more effective responses and a more thorough comprehension of the ongoing crisis. Recent studies using Twitter data have explored different aspects of Long COVID. These include the detection of self-reported symptoms from tweets \cite{Mackey2020MachineLT}, healthcare workers' perceptions in patients who are children and young people \cite{Martin2022LongCOVIDAC}, analysis of conversations \cite{Santarossa2021UnderstandingT}, and the creation of extensive collections of tweets \cite{Chen2020TrackingSM}. A recent study by Awoyemi et al. (2022) explores frequent keywords, topics, and emotions within Long COVID related tweets \cite{Awoyemi2022TwitterSA}. These studies have helped increase awareness about the impact of Long COVID on individuals and society.

However, to date, limited research has investigated the mental health of individuals with Long COVID and their interactions on social media. Nevertheless, there is a need to bridge the gap between the macro-level analysis of tweets and the micro-level analysis investigating the views and experiences of individuals. This study aims to investigate the mental health condition of Long COVID Twitter users by analyzing their tweets, examining their personal experiences, and evaluating how social media could affect their mental health and well-being.

\section{Methods}
\subsection{Data Extraction and Preprocessing}
We use SNSCRAPE\footnote[1]{https://github.com/JustAnotherArchivist/snscrape}, a widely-used web scraper, to collect tweets in English concerning post-COVID experiences with five keywords: \textit{Chronic COVID, Long COVID, Long Hauler, Myalgic Encephalomyelitis/Chronic Fatigue Syndrome (MECFS) and COVID-19, and Not Recovered}. To confirm the keywords' common usage in the context of Long COVID, we use Hashtagify\footnote[2]{https://hashtagify.me}, a hashtag tracking tool that analyzes trends in hashtag usage within tweets and has been used in previous studies \cite{TurnerMcGrievy2015TweetFH,Harris2014ArePH}. After removing duplicates, the dataset comprises 2,534,631 tweets from May 1, 2020, to January 31, 2023.

For the classification tasks, we remove web links, hashes (\#), retweets (``RT"), user mentions (@), excessive spaces, punctuations, and numbers. We convert the cleaned texts to lowercase, tokenize them into individual words, filter out stop-words, and pad the resulting sequences to ensure uniform length. Finally, we convert the preprocessed text to PyTorch tensors for use.

For topic modeling, we implement a more refined procedure. In addition to the preprocessing steps in the classification tasks above, we remove hashtags, audio/video tags, and numerical characters. We utilize Natural Language Toolkit (NLTK) to perform tokenization and lemmatization with the WordNet lemmatizer. We further refine the data by removing tokens with three or fewer characters, ultimately generating bigrams from the remaining tokens.

\subsection{Tweet Classification using pre-trained Language Models}
\subsubsection{Text Classification}
Introduced by Liu et al. \cite{Liu2019RoBERTaAR}, the Robustly Optimized BERT Pretraining Approach (RoBERTa) is distinguished from Bidirectional Encoder Representations from Transformers (BERT) by utilizing extensive data, larger batch sizes, and longer training times. Barbieri et al. \cite{Barbieri2020TweetEvalUB} has demonstrated the efficacy of RoBERTa-based models in the text classification of tweets. 

After manually labeling data, we employ the RoBERTa model to perform a hierarchical text classification comprised of three binary tasks. First, we split our data into two categories based on whether the tweet includes self-reported Long COVID symptoms. Second, we identify if the self-reported Long COVID tweet includes any instances of self-reported mental health issues. Lastly, for the self-reported mental health tweets, we categorize them as either implicit or explicit based on the presence of mental health statements (anonymized examples in Table \ref{tab1}). For simplicity, we use the following abbreviations for each category: (1) Non-Long COVID Tweets (NC), (2) Long COVID Tweets without Mental Health Content (LC), (3) Long COVID Tweets with Implicit Mental Health Statements (LC-I), and (4) Long COVID Tweets with Explicit Mental Health Statements (LC-E). 

\vspace{-15pt}

\begin{table}[ht]
\centering
    \caption{Examples of Tweets in Four Categories}\label{tab1}
    \label{crouch}
    \begin{tabular}{  p{1cm} p{0.5cm} p{10cm} }
        \toprule
\textbf{Category} && \textbf{Example}  \\\midrule
NC &
& Today is \#LongCovidAwarenessDay.\\\hline
LC &       
& @USER I've been living with \#LongCovid for 21 months. This was huge for me. \\\hline
LC-I &       
& I do take Cymbalta for  nerve/muscle/bone pain due to \#LongCovid. This med can help  many different symptoms..anxiety, sleep issues, nerve issues. \\\hline
LC-E &        
& @USER @USER I'm already on meds for depression due to Long Covid...Shoot me now.  \\
        \bottomrule
    \end{tabular}
    \vspace{-20pt}
\end{table}

\subsubsection{Emotion Classification}
Emotions are vital in human experiences and psychiatric illnesses \cite{Ehrenreich2007TheRO}. We follow Butt et al. \cite{Butt2021WhatGO} to employ a distilled version of the RoBERTa model (DistilRoBERTa) to detect the presence of Ekman's six basic emotions \cite{Ekman1992AnAF} (Anger, Disgust, Fear, Joy, Sadness, and Surprise) within tweets. 

\subsection{Topic Modeling with the latent Dirichlet allocation (LDA)}
Topic modeling is an automatic, unsupervised machine learning technique widely used to detect latent topics for a collection of documents. We apply LDA \cite{Blei2001LatentDA} since it is one of the most widely used models for topic modeling and tends to produce more coherent topics on tweets compared to other models \cite{Culmer2021ExaminingLA}. We specify a range for the topic number to determine the optimal number. We calculate the coherence scores for each topic number using the ``UMass" measure. A lower ``UMass" score indicates higher readability of the results. Finally, we select the number of topics corresponding to the lowest coherence score. 

\subsection{Spearman's Rank Correlation Coefficient}
We use Spearman's rank correlation coefficient or Spearman's $\rho$, denoted as $r_s$, to investigate whether one category's emotions co-variate with another's. It is a statistical measure that utilizes a monotonic function to determine relations between variables. Unlike other methods, such as Pearson's correlation, it does not rely on the normality of data and is appropriate for continuous and discrete ordinal variables. Thus, it provides a suitable instrument for assessing relations between time series data like emotion variables. Spearman's $\rho$ ranges from -1 to 1, where -1 indicates a perfect negative correlation, 0 indicates no correlation, and 1 indicates a perfect positive correlation between the two variables.

\section{Results}
\vspace{-10pt} 

\subsection{RQ1. What are the distributions of emotions of different categories, and how do these emotions change over time?}

\vspace{-15pt} 

\begin{figure}[ht]
\centering
\includegraphics[width=0.8\textwidth]{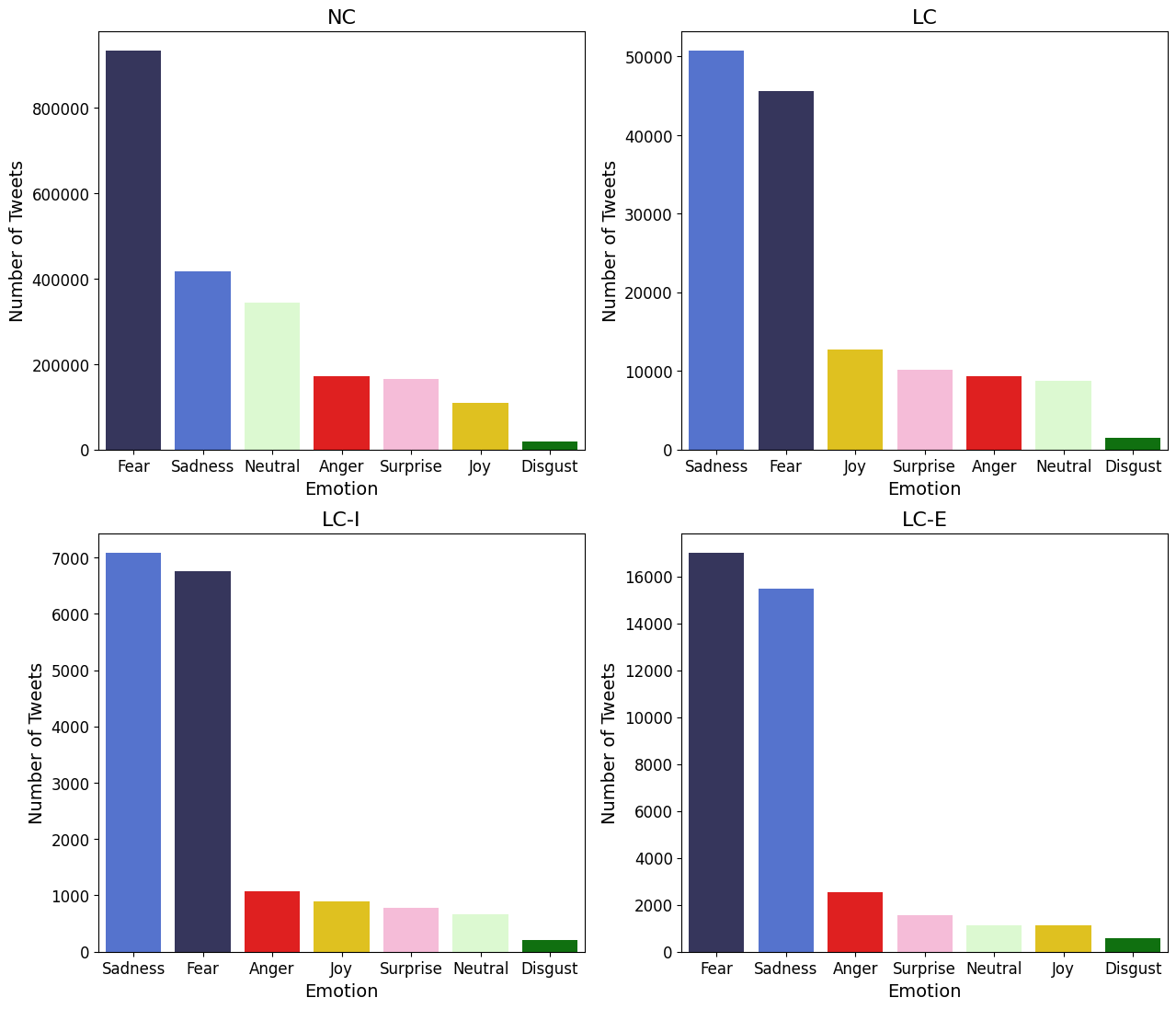}
\caption{Distributions of Ekman's Six Basic Emotions by Categories} \label{fig2}
\end{figure}

\vspace{-10pt} 

Overall, we find that Fear is the dominant emotion among all tweets, accounting for 42.57\% of all tweets, followed by Sadness (20.82\%), Neutral (15.03\%), Anger (7.85\%), Surprise (7.56\%), Joy (5.27\%), and Disgust (0.88\%). 

The emotion distributions by categories are shown in Fig. \ref{fig2}. Notably, a significant increase in Sadness can be observed among all LC, LC-I, and LC-E categories. Specifically, the percentages of Sadness are up by 15.7\% (LC), 19.7\% (LC-I), and 18.4\% (LC-E) compared to the overall emotion distribution. While the percentage of Sadness is slightly higher than Fear in LC and LC-I, Fear is the primary emotion expressed in tweets in LC-E.

\begin{figure}[ht]
\centering
\includegraphics[width=0.8\textwidth]{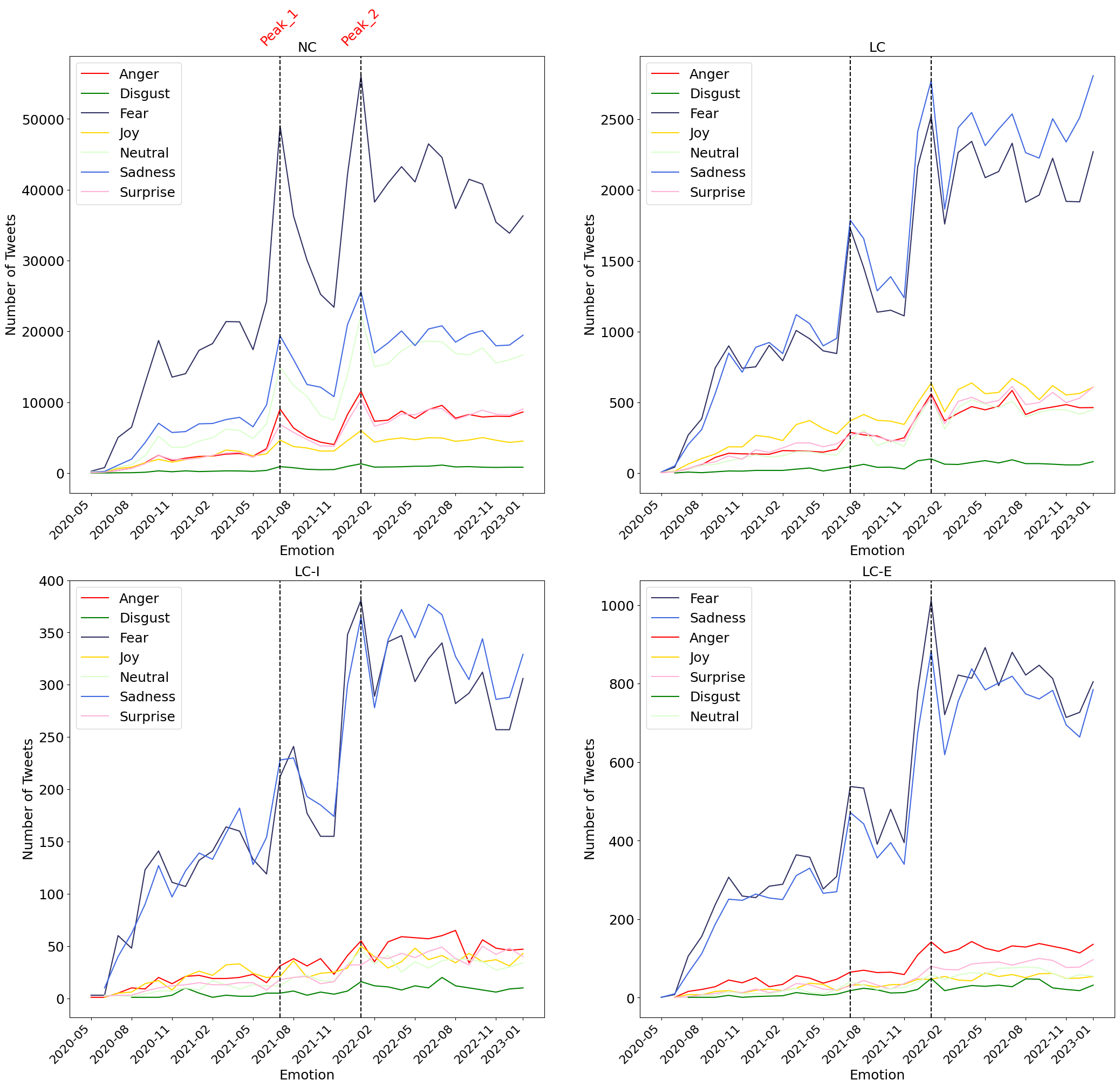}
\caption{Monthly Frequency of Ekman's Six Basic Emotions by Categories} \label{fig3}
\end{figure}

The monthly frequency of the number of tweets of Ekman's emotions by categories in Fig. \ref{fig3} shows that Fear and Sadness dominate our study time period. In the monthly trends of emotions in LC, LC-I, and LC-E, we notice two spikes of Fear and Sadness in July 2021 and January 2022. This pattern persists in the monthly trend of emotions in NC and reveals more emotions, including Neutral, Joy, and Surprise.

\subsection{RQ2. What are the correlations between the emotions of different categories?}
We use Spearman's $\rho$ to calculate the $r_s$ coefficients. We set the weak correlation interval from 0.00 to 0.30, the moderate correlation interval from 0.30 to 0.70, and the strong correlation interval from 0.70 to 1.00, according to \cite{iversen2012statistics}.

The correlation between tweets of the three categories (LC, LC-I, and LC-E) and tweets of NC reveals that Fear and Sadness expressed in NC are mostly positively correlated (moderate to strong) with those of the three categories.

We then calculate the $r_s$ coefficients among NC, LC, LC-I, and LC-E. Between NC and LC, Fear and Sadness are mostly correlated. They show a strong positive correlation, mainly in July and December 2021, January and February 2022. Looking at the correlations between Fear and Sadness in NC and LC-I among 33 months, only 33.3\% of the time are correlated. These two negative emotions show a moderate positive correlation in August 2021 and a strong positive correlation in December 2021. If we only consider the correlation between Fear and Sadness, 61\% of them are correlated between NC and LC-E. A moderate positive correlation exists in July 2021 and August 2021. In December 2021 and January 2022, the correlation between these two negative emotions is strong.

By analyzing correlation results, we find that the emotional distribution of LC-I is less responsive to NC's emotional swing than LC and LC-E. Considering the change of correlations over time, we find two periods (July - August 2021 and December 2021 - January 2022) that show a high positive correlation corresponding to the outburst of the Delta and Omicron variants, respectively. These variants may significantly impact the emotional swing of both NC and LC.

\subsection{RQ3. What are the prevalent topics in different categories?}
We apply two LDA models on NC in July 2021 and NC in January 2022 since these are the peak periods we identified in RQ1, and three LDA models for LC, LC-I, and LC-E. We then group all 102 topics from five LDA models into 17 overarching themes, see Table \ref{tab2}.

\begin{table}[ht]
\centering
\vspace{-15pt}
\caption{17 Overarching Themes}\label{tab2}
    \begin{tabular}{p{5cm} p{0.25cm} p{6.5cm}}
    \hline
        1. mask wearing && 2. positive attitude and appreciation for life\\ \hline
        3. medical assistance/research && 4. fatigue\\ \hline
        5. vaccine benefit and risk concerns && 6. clinical advice/diagnosis\\ \hline
        7. concern on children && 8. discussions about Omicron\\ \hline
        9. vaccine advocacy && 10. diseases potentially related to COVID-19\\ \hline
        11. physical pain && 12. neurological disorders\\ \hline
        13. suicidal tendency && 14. anxiety and restlessness about life\\ \hline
        15. smell/taste disorders && 16. insomnia/lack of sleep\\ \hline
        17. cognitive/memory decline\\ \hline
        
    \end{tabular}
\end{table}
\vspace{-15pt} 

For NC in July 2021, Theme 7 is the most frequent, accounting for 60.01\% among total tweets. RQ1 shows that the Delta variant was the dominant variant in July 2021. The Delta variant quickly overtook the Alpha variant through multiple and repeated infections. It was already translating into increased hospitalizations and deaths according to the \textit{Statement by Dr Hans Henri P. Kluge}\footnote[3]{https://www.who.int/europe/news/item/01-07-2021-covid-19-the-stakes-are-still-high}. We can infer that the Delta variant showed a higher risk, so parents started to worry about their kids in school. For NC in January 2022, there are two highly frequent themes: Theme 8 (48.33\%), Theme 3 (18.27\%), and three themes related to symptoms of Omicron: Theme 4 (15.44\%), Theme17 (8.55\%) and Theme15 (1.7\%). From \textit{The Omicron variant: sorting fact from myth}\footnote[4]{https://www.who.int/europe/news/item/19-01-2022-the-omicron-variant-sorting-fact-from-myth}, we learn that the medical requirement suddenly increased, and the weak who had some underlying diseases died because of the Omicron variant. From \textit{WHO's news: Living with COVID-19 – 2 years on}\footnote[5]{https://www.who.int/europe/news/item/27-01-2022-living-with-covid-19-2-years-on}, post-COVID conditions happened on many people who have infected by the Omicron variant. These two pieces of news validate our LDA results. Furthermore, because the infection rate of Omicron was faster than ever, people discussed it continuously during that period.

For LC, we find that the predominant topics of LC are different. Theme 2 (37.38\%), 11 (23.33\%), 4 (21.76\%), 16 (6.16\%), and 15 (1.89\%)  represent most of the content this category discussed. Therefore, some people whose tweets have been classified in this category discussed more symptoms and may have recovered quickly from COVID-19 and have taken a positive attitude toward life. For LC-I topics, we find that more tweets are related to sequelae corresponding to Themes 4, 11, 15, and 16. Moreover, 0.76\% of people started to post content about Theme 12. There are 8.14\% of the tweets posting suicidal talk corresponding to Theme 14. For LC-E, most tweets may have suffered from the sequelae of COVID-19 since the percentage of those discussions is higher than LC-I. 28.95\% of tweets are beginning to complain about Theme 17. The percentage of Theme 6, Theme 13, and Theme 14 are 10.82, 1.71, and 18.45, respectively. We conclude that people in this category suffer from more serious sequelae, like cognitive decline, and have a negative attitude toward life; even a small number have extremely negative emotions.

\section{Discussions}
This study examines tweets about Long COVID between May 2020 and January 2023. Most tweets in our dataset do not contain self-reported Long COVID information. Additionally, around 30\% of the tweets of the three categories (LC, LC-I, and LC-E) discuss mental health issues.

Results of the emotion analysis demonstrate that Fear and Sadness are prevalent in over 50\% of the tweets in our dataset and persist throughout the entire study period. These negative emotions peaked in terms of the number of tweets in July 2021 and January 2022, coinciding with the emergence of new COVID-19 variants and the strain on medical resources. Our correlation analysis reveals that the emotions expressed by four categories exhibit different patterns. While negative emotions such as Fear and Sadness in tweets by LC, LC-I, and LC-E tend to vary along with those in the tweets by NC, the correlation between the expression of other emotions such as Joy, Surprise, and Neutral is only trivial. 

NC tweets mainly discuss children's health, COVID infections, medical instructions, and research. LC tweets contain more content expressing positive attitudes and gratitude towards life, whereas LC-I and LC-E tweets often discuss COVID symptoms and sequelae. In particular, tweets in LC-E are not limited to mental health issues such as anxiety, depression, autism, and others but also include discussions of suicidal thoughts.

Our findings reveal that public health emergencies have the potential to amplify the volume and intensity of discussions taking place on social media platforms, which may have an impact on negative emotions expressed by those who report experiencing Long COVID symptoms. The different emotional variations between the four categories suggest that targeted interventions may be necessary for different groups' emotional needs. Furthermore, the topics discussed in tweets by LC, LC-I, and LC-E highlight the need for increased awareness and support for individuals experiencing COVID symptoms and sequelae, particularly regarding mental health. 

While our study provides valuable insights into individuals' mental health and emotional experiences with Long COVID, limitations exist. We rely on Twitter data from keyword searches and self-reported Long COVID Twitter users. This may only capture some Long COVID-related discussions on social media, and there may be a bias toward more active individuals on Twitter. We only analyze Twitter data, which may not represent the experiences and emotions of individuals with Long COVID in the general population. The study also does not consider demographic factors such as age, sex, and location, which may impact the emotions and experiences of individuals with Long COVID. We focus on emotions and topics related to Long COVID and mental health. Other factors, such as social support, financial impact, and access to healthcare, may also contribute to the experiences of individuals with Long COVID and are not analyzed in this study. In the future, we plan to incorporate the abovementioned factors and data from patient support groups and other sources to gain a broader perspective on the experiences of individuals with Long COVID.

\section{Conclusions}
In conclusion, this study comprehensively analyzes the emotions and topics discussed on Twitter about Long COVID from May 2020 to January 2023. The findings highlight the prevalence of Fear and Sadness in Long COVID tweets and the impact of public health emergencies on negative emotions expressed in social media discussions. The methodology employed in this study can inform future analyses of social media trends for public health emergencies. Overall, this study contributes to a better understanding individuals' emotional and mental well-being with Long COVID and provides valuable insights for targeted interventions and support.


\bibliographystyle{splncs04}
\bibliography{resources/citation}

\end{document}